\documentclass[fleqn]{2020SCGE}
\usepackage{multirow}
\usepackage{siunitx}
\usepackage{hyperref}

\usepackage{color}

\begin{document}

\ensubject{subject}

\ArticleType{Article}
\SpecialTopic{SPECIAL TOPIC: }
\Year{2023}
\Month{xxx}
\Vol{xx}
\No{x}
\DOI{xx}
\ArtNo{000000}
\ReceiveDate{December 13, 2023}
\AcceptDate{February 26, 2023}
\AuthorMark{Dengke Zhou}

\AuthorCitation{Dengke Zhou, Pei Wang, Di Li, Jianhua Fang, Chenchen Miao, Paulo C. C. Freire, Lei Zhang, Dandan Zhang, Huaxi Chen, Yi Feng, Yifan Xiao, Jintao Xie, Xu Zhang, Chenwu Jin, Han Wang, Yinan Ke, Xuerong Guo, Rushuang Zhao, Chenhui Niu, Weiwei Zhu, Mengyao Xue, Yabiao Wang, Jiafu Wu, Zhenye Gan, Zhongyi Sun, Chengjie Wang, Jie Zhang, Junshuo Zhang, Jinhuang Cao, Wanjin Lu}

\title{A discovery of Two Slow Pulsars with FAST: ``Ronin" from the Globular Cluster M15}

\author[1]{\\Dengke Zhou}{}
\author[2,3]{Pei Wang\footnote{Corresponding author. Email: wangpei@nao.cas.cn}}{}
\author[2,4,5]{Di Li\footnote{Corresponding author. Email: dili@nao.cas.cn}}{}
\author[1]{Jianhua Fang}{}
\author[1]{Chenchen Miao}{}
\author[6]{Paulo C. C. Freire}{}
\author[2,7]{Lei Zhang}{}
\author[8,9]{\\Dandan Zhang}{}
\author[1]{Huaxi Chen}{}
\author[1]{Yi Feng}{}
\author[10]{Yifan Xiao}{}
\author[1]{Jintao Xie}{}
\author[1]{Xu Zhang}{}
\author[1]{Chenwu Jin}{}
\author[1]{Han Wang}{}
\author[1]{\\Yinan Ke}{}
\author[1]{Xuerong Guo}{}
\author[11]{Rushuang Zhao}{}
\author[12]{Chenhui Niu}{}
\author[2,3]{Weiwei Zhu}{}
\author[2]{Mengyao Xue}{}
\author[13]{Yabiao Wang}{}
\author[13]{\\Jiafu Wu}{}
\author[13]{Zhenye Gan}{}
\author[13]{Zhongyi Sun}{}
\author[13]{Chengjie Wang}{}
\author[10]{Jie Zhang}{}
\author[2,4]{Junshuo Zhang}{}
\author[2,4]{\\Jinhuang Cao}{}
\author[2,4]{Wanjin Lu}{}
\address[1]{Research Center for Astronomical Computing, Zhejiang Laboratory, Hangzhou {\rm 311121}, China}
\address[2]{National Astronomical Observatories, Chinese Academy of Sciences, Beijing {\rm 100101}, China}
\address[3]{Institute for Frontiers in Astronomy and Astrophysics, Beijing Normal University, Beijing {\rm 102206}, China}
\address[4]{University of Chinese Academy of Sciences, Beijing {\rm 100049}, China}
\address[5]{New Cornerstone Science Laboratory, Shenzhen 518054, China}
\address[6]{Max Planck Institut f\"ur Radioastronomie, Bonn {\rm 53121}, Germany}
\address[7]{Centre for Astrophysics and Supercomputing, Swinburne University of Technology, Hawthorn {\rm 3122}, Australia}
\address[8]{School of Mathematical Sciences, School of Physics and Electronic Sciences, Guizhou Normal University, Guiyang {\rm 550001}, China}
\address[9]{Guizhou Provincial Key Laboratory of Radio Astronomy and Data Processing, Guiyang {\rm 550001}, China}
\address[10]{School of Physics and Electronic Engineering, Qilu Normal University, Jinan {\rm 250200}, China}
\address[11]{School of Physics and Electronic Science, Guizhou Normal University, Guiyang {\rm 550001}, China}
\address[12]{College of Physical Science and Technology, Central China Normal University, Wuhan {\rm 430079}, China}
\address[13]{Tencent Youtu Lab, Shanghai {\rm 201103}, China}

\abstract{Globular clusters harbor numerous millisecond pulsars, but long-period pulsars ($P \gtrsim 100$ ms) are rarely found. In this study, we employed a fast folding algorithm to analyze observational data from multiple globular clusters obtained by the Five-hundred-meter Aperture Spherical radio Telescope (FAST), aiming to detect the existence of long-period pulsars. We estimated the impact of the median filtering algorithm in eliminating red noise on the minimum detectable flux density ($S_{\rm min}$) of pulsars. Subsequently, we successfully discovered two isolated long-period pulsars in M15 with periods approximately equal to 1.928451 seconds and 3.960716 seconds, respectively. On the $P-\dot{P}$ diagram, both pulsars are positioned below the spin-up line, suggesting a possible history of partial recycling in X-ray binary systems disrupted by dynamical encounters later on. According to timing results, these two pulsars exhibit remarkably strong magnetic fields. If the magnetic fields were weakened during the accretion process, then a short duration of accretion might explain the strong magnetic fields of these pulsars.}



\maketitle


\begin{multicols}{2}
\section{Introduction}
Globular clusters, characterized by their high stellar density, are environments where stellar collisions and interactions are prevalent throughout their evolutionary history, giving rise to a plethora of pulsar binaries, including millisecond pulsars (MSPs) \cite{MSP_1975ApJ...199L.143C, MSP1_1991PhR...203....1B}. Consequently, globular clusters have long been recognized as crucial birthplaces for MSPs. To date, hundreds of MSPs have been discovered in globular clusters, providing essential samples for studying the evolutionary mechanisms of these pulsars.

In contrast, long-period pulsars ($P \gtrsim 100$ ms) are notably rare within globular clusters. Thus far, only two instances of long-period pulsars with periods on the order of seconds have been discovered in globular clusters. One pulsar has a period of approximately 1 second and resides in a binary system with a companion mass of around 0.1 solar masses \cite{long1_1993Natur.361...47L}. The other pulsar, with a period of about 2.5 seconds, is an isolated pulsar, and its affiliation with a globular cluster is yet to be confirmed \cite{long2_2022MNRAS.513.2292A}.

The scarcity of long-period pulsars underscores their significance in unraveling the evolutionary pathways of pulsars within globular clusters. The presence of long-period pulsars may imply the existence of distinct dynamical and stellar evolution processes in globular clusters, divergent from the formation mechanisms of MSPs, indicating an alternative pulsar evolutionary path. The rarity of these long-period pulsars emphasizes the urgent need for in-depth investigations. Through more extensive searches and studies of this class of pulsars, we anticipate uncovering the subtle and complex structures within globular clusters, providing more accurate clues to interpret the past and future of these ancient stellar systems.

The FAST telescope, as the current world's largest single-dish telescope, possesses extremely high sensitivity in pulsar searches, making the exploration of new pulsars one of its primary objectives \cite{CRAFTS_main_8331324, CRAFTS_FASTFP_2019SCPMA..6259508Q, CRAFTS_pan_2020ApJ...892L...6P, CRAFTS_earlyPsr_2021MNRAS.508..300C, Lei1_2019ApJ...877...55Z}. In particular, FAST has conducted frequent and prolonged observations of globular clusters, providing ample data for the search of long-period pulsars in globular clusters \cite{CRAFTS_pan_2020ApJ...892L...6P, GC_pan_2021ApJ...915L..28P, lei2_Zhang_2023}.

Challenges in searching for long-period pulsars persist, primarily due to the impact of red noise introduced by prolonged integration time (e.g., references \cite{simulation_FFTg100_2015ApJ...812...81L,longPsearch2_parent_kaspi_ransom_patel_krasteva_2017, longPsearch_2022ApJ...934..138S}). Additionally, the issue of pulse nulling significantly affects the search for periodic signals based on Fourier transform techniques \cite{nulling_1970Natur.228...42B, FFA_nulling_2023ApJ...954..160S}. In this regard, the fast folding algorithm (FFA) has demonstrated higher sensitivity in searching for periodic signals compared to Fourier transform-based search algorithms \cite{FFA_2020MNRAS.497.4654M}.

In this paper, we aim to utilize FFA to search for long-period pulsars in FAST observations of globular cluster data. We hope to unveil evolutionary branches distinct from MSPs within these ancient celestial systems. The structure of this paper is as follows: Section \ref{sec:tech} details the search techniques employed, including algorithmic descriptions and parameter settings. Section \ref{sec:discoveries} presents our search outcomes. Section \ref{sec:discussion} discusses the two discovered pulsars. Finally, Section \ref{sec:summary} provides a summary of the entire paper.

\section{Observation and Data Reduction}\label{sec:tech}
\begin{table*}
	\centering
	\caption{Information of the globular clusters searched in this paper.}  
	\label{tab:data}
	\small
	\begin{threeparttable}
                   \begin{tabular}{llccccc}
				\toprule
				GC Name & R.A. & Decl. & Observation & Observation  & DM & Sensitivity\tnote{*}\\
				      & (J2000) & (J2000) & Date (UT) & Length (hr) & ($\rm pc\ cm^{-3}$) & ($\mu {\rm Jy}$)\\
				\midrule
				M53(NGC 5024) & 13:12:55 & +18:10:05.4 & 2019/11/30 & 5.00 & 24-26.2 & 1.3$\sim$12.4\\
				NGC5053 & 13:16:27 & +17:42:00.9 & 2019/12/25 & 2.33 & 20.23 (YMW16) & 2.0$\sim$18.3\\
				M3(NGC 5272) & 13:42:12 & +28:22:38.2 & 2019/12/14 & 4.50 & 26.34-26.541 & 1.1$\sim$10.4\\
				NGC 5466 & 14:05:27 & +28:32:04.0 & 2019/12/23 & 5.00 & 20.721(YMW16) & 1.4$\sim$12.6\\
				NGC 5634 & 14:29:37 & -05:58:35.1 & 2020/01/05 & 1.00 & 28.291(YMW16) & 3.5$\sim$33.5\\
				M5(NGC 5904) & 15:18:33 & +02:04:51.7 & 2019/12/11 & 4.00 & 29.3-30.8 & 1.7$\sim$16.4\\
				M5(NGC 5904) & 15:18:33 & +02:04:51.7 & 2020/01/06 & 4.00 & 29.3-30.8 & 1.5$\sim$14.6\\
				NGC 6229 & 16:46:59 & +47:31:39.9 & 2020/01/03 & 3.83 & 31.109(YMW16) & 1.7$\sim$16.0\\
				NGC 6229 & 16:46:59 & +47:31:39.9 & 2020/01/16 & 3.83 & 31.109(YMW16) & 1.5$\sim$14.2\\
				M10(NGC 6254) & 16:57:09 & -04:06:01.1 & 2020/01/05 & 3.00 & 43.355-43.9 & 1.9$\sim$18.0\\
				M92(NGC 6341) & 17:17:07 & +43:08:09.4 & 2020/01/08 & 0.50 & 35.45 & 5.9$\sim$57.0\\
				M92(NGC 6341) & 17:17:07 & +43:08:09.4 & 2020/01/09 & 0.50 & 35.45 & 6.5$\sim$59.4\\
				M92(NGC 6341) & 17:17:07 & +43:08:09.4 & 2020/01/13 & 0.50 & 35.45 & 6.2$\sim$57.3\\
				M14(NGC 6402) & 17:37:36 & -03:14:45.3 & 2020/01/08 & 1.00 & 78.8-80.4 & 4.3$\sim$40.6\\
				M14(NGC 6402) & 17:37:36 & -03:14:45.3 & 2020/01/09 & 1.00 & 78.8-80.4 & 4.0$\sim$39.3\\
				M14(NGC 6402) & 17:37:36 & -03:14:45.3 & 2020/01/13 & 1.00 & 78.8-80.4 & 3.9$\sim$38.9\\
				NGC 6426 & 17:44:55 & +03:10:12.5 & 2019/11/29 & 4.00 & 87.767(YMW16) & 1.5$\sim$14.8\\
				NGC 6749 & 19:05:15 & +01:54:03 & 2019/10/26 & 1.00 & 192-193.692 & 3.7$\sim$34.4\\
				NGC 6749 & 19:05:15 & +01:54:03 & 2019/12/10 & 3.00 & 192-193.692 & 1.8$\sim$15.7\\
				M56(NGC 6779) & 19:16:36 & +30:11:00.5 & 2019/12/01 & 5.00 & 101.359(YMW16) & 1.3$\sim$12.7\\
				M71(NGC 6838) & 19:53:46 & +18:46:45.1 & 2019/12/12 & 5.00 & 113.1-119.038 & 1.4$\sim$13.5\\
				M72(NGC 6981) & 20:53:28 & -12:32:14.3 & 2020/01/06 & 1.00 & 40.472(YMW16) & 8.6$\sim$83.4\\
				M15(NGC 7078) & 21:29:58 & +12:10:01.2 & 2019/11/09 & 1.33 & 65.52-67.69 & 3.0$\sim$26.7\\
				M2(NGC 7089) & 21:33:27 & -00:49:23.7 & 2019/11/16 & 2.00 & 43.3-44.1 & 2.2$\sim$20.0\\
				M2(NGC 7089) & 21:33:27 & -00:49:23.7 & 2019/11/17 & 2.00 & 43.3-44.1 & 2.2$\sim$21.4\\
				M2(NGC 7089) & 21:33:27 & -00:49:23.7 & 2019/11/18 & 2.00 & 43.3-44.1 & 2.2$\sim$20.9\\
				M2(NGC 7089) & 21:33:27 & -00:49:23.7 & 2020/01/04 & 2.00 & 43.3-44.1 & 2.3$\sim$22.0\\
				\bottomrule
			\end{tabular}
		\begin{tablenotes}
			\item[*] The Sensitivity is the minimum detectable flux density assuming a duty cycle of 10\%, SNR of 6, spin periods of 1-100 seconds, and integration time for the corresponding observational data using FAST L-band receiver (see in Section \ref{sec:red_smin}).
		\end{tablenotes}
	\end{threeparttable}
\end{table*}

We conducted a search on publicly available partial globular cluster data from FAST (1.0$\sim$1.5 GHz), and the information for these globular clusters is listed in Table \ref{tab:data}. The observation times for these clusters mostly range from one to a few hours, providing us with sufficient signal-to-noise ratio (SNR) for detecting faint pulsars, especially long-period pulsars.

\subsection{FFA Algorithm and Search Process}\label{sec:FFA}
The FFA differs from the commonly used fast Fourier transform (FFT) for pulsar periodic signal searches. This algorithm folds the time series containing pulsar signals at regular intervals. During the folding process, the signal is enhanced more than the noise, resulting in an increased signal-to-noise ratio. Ultimately, the signal's most significant pulse profile emerges when folded at the true period. By testing different periods, obtaining the significance of the pulsar profile as a function of the test period allows the identification of the optimal period corresponding to the maximum significance. This method exhibits significant advantages over FFT in searching for weak signals, signals with pulse nulling, and long-period signals \cite{simulation_FFTg100_2015ApJ...812...81L, Pg2_2009ApJ...702..692K,FFA_2020MNRAS.497.4654M, FFA_nulling_2023ApJ...954..160S}. However, due to the relatively high computational cost compared to FFT, this algorithm has been used less frequently in recent years in the field of radio pulsar searches.

Recently, reference \cite{FFA_2020MNRAS.497.4654M} proposed that the sensitivity of FFA in searching for pulsars at any period surpasses that of FFT, especially when searching for long-period pulsars. This prompts us to reconsider applying the algorithm to search for long-period pulsars. Especially for target sources like globular clusters, most of which have known dispersion measure (DM), it can help reduce the number of DMs to search, making it feasible to apply FFA in pulsar searches. Additionally, FAST has relatively long observation durations for these sources (from one hour to several hours), increasing our confidence in the detectability of these pulsars. However, due to fluctuations in telescope system temperature, gain, and celestial background, strong red noise is introduced into the data. This poses challenges for the search of long-period pulsars, especially those that are faint and dim. We will assess this issue in Section \ref{sec:red_smin}.

FFA folds the time series according to a presumed period, obtaining an averaged time series within one period. Assuming the whitened (red noise removal) and normalized (subtracting the mean and dividing by the standard deviation) time series after folding is represented by vector $\mathbf{x}$, the folded signal profile is represented by a vector $a\mathbf{s}$, and the folded background noise is represented by vector $\mathbf{w}$, the equation is given by
\begin{equation}
	\mathbf{x}=a\mathbf{s}+\mathbf{w}.
\end{equation}
Here, $\mathbf{s}$ represents the normalized pulse signal, and $\mathbf{a}$ represents the scaling factor for the normalized pulse signal. $\mathbf{s}$ satisfies $\sum\limits_{i=1}^n s_i^2=\mathbf{s}\cdot \mathbf{s}=1$ and $\sum\limits_{i=1}^n s_i=0$. If $a>0$, it indicates the presence of a pulsar signal in $\mathbf{x}$. To verify the presence of a pulsar signal in $\mathbf{x}$, a direct method is to use a known pulsar profile template $\mathbf{s}$ and perform a dot product with $\mathbf{x}$. The resulting value is called the Z-statistic, given by
\begin{equation}\label{eq: SNRz}
	Z = \mathbf{x}\cdot \mathbf{s} = a+\mathbf{w}\cdot\mathbf{s}=a+\sum\limits_{i=1}^ns_iw_i.
\end{equation}
As $\mathbf{x}$ is whitened and normalized, its mean value is 0. Assuming that the contribution of the pulsar signal $a\mathbf{s}$ to determining the overall profile mean is much smaller than the contribution of the noise $\mathbf{w}$, then $w_i\sim \mathcal{N}(0,1)$. Thus, $\sum\limits_{i=1}^ns_iw_i\sim \mathcal{N}(0, \sum\limits_{i=1}^n s_i^2)$, considering $\sum\limits_{i=1}^n s_i^2=1$. Consequently, $Z\sim \mathcal{N}(a, 1)$. In practical searches, the actual pulsar profile template $\mathbf{s}$ is unknown. However, a series of normalized box cars with varying widths and phases can be used to progressively approximate the pulsar profile template $\mathbf{s}$. Reference \cite{FFA_2020MNRAS.497.4654M} suggests continuing to refer to Z as the detection SNR, and SNR can be used to filter candidate signals.
\begin{figure*}
	\centering
	\includegraphics[scale=0.55]{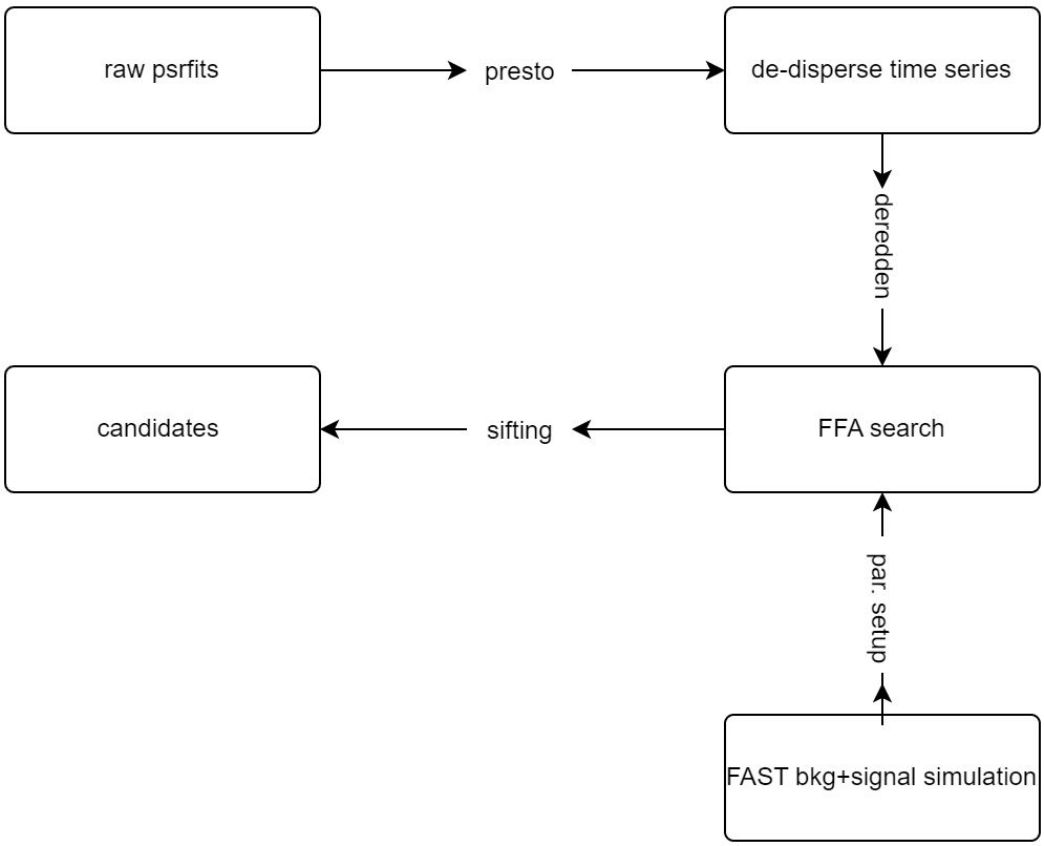}
	\caption{The flowchart of searching for long-period pulsars in globular clusters in this paper.}
	\label{fig:pipline}
\end{figure*}

Our search process is as follows. Initially, \emph{presto}\footnote{\url{https://github.com/scottransom/presto}} is utilized to mitigate radio frequency interference (RFI) in the data, followed by batch de-dispersion processing to obtain a large number of time series corresponding to different DM. In these globular clusters, most of them have detected several pulsars, enabling us to concentrate the de-dispersion range around a good value. If no pulsars have been detected in a specific globular cluster, making it impossible to determine the DM, we use the YMW16 model to estimate its DM \cite{YMW16_2017ApJ...835...29Y}. Additionally, we appropriately downsampled the data in time to reduce the data volume and improve the efficiency of searching for long-period pulsars. After obtaining the time series, it is necessary to remove red noise from the data, as our main focus is on searching for long-period pulsars. The removal of red noise and FFA search are both performed using \emph{riptide}\footnote{\url{https://github.com/v-morello/riptide}}. As revealed in the simulations in the subsequent section, although the red noise in FAST data does not significantly impact the minimum detectable flux density of pulsars, the process of removing red noise weakens the amplitude of the signal itself, thus affecting the minimum detectable flux density. We will discuss this in detail in the Section \ref{sec:red_smin}. Fig. \ref{fig:pipline} shows the flowchart of our search for long-period pulsars in globular clusters.

Additionally, it is worth noting that in this paper, we have exclusively conducted optimized searches for pulse period and width, without considering the Doppler phase shift induced by binary motion. However, this does not imply that our search is limited to isolated pulsars. Strictly speaking, this work encompasses the search for some binary systems, but it does not specifically address the Doppler effect correction for binary system to enhance detection rates. For binary system with elliptical orbit, the phase $\Phi(t)$ of the signal follows the following equations \cite{2001PhRvD..63l2001D}\\
\begin{align}
\Phi(t) &= 2\pi f_0t + \phi_D(t), \\
\begin{split}
\phi_D(t) &= -\frac{2\pi f_0a\sin\epsilon}{c} \Big[\cos\psi\cos E(t) \\
&\quad + \sin\psi\sqrt{1-e^2}\sin E(t) \Big],
\end{split} \label{eq:D} \\
E(t) - e\sin E(t) &= \Omega + \alpha,
\end{align}
where $f_0$ is the pulsar's spin frequency, and $\phi_D(t)$ is the pulse phase shift caused by the Doppler effect due to the binary motion. From equation (\ref{eq:D}), it can be seen that pulsars with longer periods (lower frequencies) experience smaller phase shifts due to the Doppler effect. In this paper, our primary focus is on the search for long-period pulsars (0.1 seconds to 100 seconds). The phase shift caused by the binary motion is therefore limited in its impact on our search target. A detailed discussion of the specific effects is beyond the scope of this paper; however, we plan to address this issue in future searches.

\subsection{Calculating Sensitivity in the Presence of Red Noise}\label{sec:red_smin}
In pulsar searches, to estimate the telescope's sensitivity to pulsars, the radiometer equation proposed by references \cite{radiometer_1985ApJ...294L..25D, handbookofPsr_2012hpa..book.....L} under the assumption of white noise can be used to estimate the average minimum detectable flux density, given by
\begin{equation}\label{eq:ro}
	S_{\rm min}=\frac{\beta(T_{\rm sys}+T_{\rm sky}){\rm SNR}}{{\epsilon G\sqrt{n_{\rm p}t_{\rm obs}\Delta f}}}\sqrt{\frac{W}{P-W}},
\end{equation}
where $P$ is the period of the pulsar signal, $\beta$ represents data digitization losses, $T_{\rm sys}+T_{\rm sky}$ is the sum of the telescope system temperature and the sky background temperature, $G$ is the telescope gain, SNR denotes the signal-to-noise ratio of the pulsar signal, $n_{\rm p}$ is the number of polarization channels for data merging, $\Delta f$ is the frequency bandwidth for data merging, and $t_{\rm obs}$ is the observation duration. The original formula does not include the parameter $\epsilon$, but according to the suggestion by reference \cite{FFA_2020MNRAS.497.4654M}, the algorithm search efficiency $\epsilon$ needs to be introduced to adjust the original radiometer equation. $W$ refers to the width of the observed pulsar signal, encompassing both the intrinsic width of the pulsar signal and the broadening due to sampling time, scattering, and channel dispersion. Using $W_{0}$ to represent the intrinsic width of the pulsar signal and $t_{\rm sample}$, $t_{\rm DM}$, and $t_{\rm scatt}$ for the sampling time, interstellar medium dispersion in the frequency channel, and interstellar scattering-induced pulse broadening, respectively, we obtain the total width
\begin{equation}
	W = \sqrt{W_0^2+t_{\rm sample}^2+t_{\rm DM}^2+t_{\rm scatt}^2},
\end{equation}
where\begin{multline}
	\log_{10}[t_{\rm scatt}(ms)] = -6.46+0.154\log_{10}({\rm DM})+\\
	1.07[\log_{10}({\rm DM})^2-3.86\log_{10}(f({\rm GHz}))]
\end{multline} \cite{tscatt_2004ApJ...605..759B}. For FAST, the typical minimum detectable flux density of pulsars can be calculated by taking $T_{\rm sys}+T_{\rm sky}=25\ \rm K$, $G=16 \ \rm K\ Jy^{-1}$, $n_{\rm p}=2$, $\Delta f=300\times 10^6 {\rm Hz}$ (considering the proportion of remaining frequency channels after excluding RFI, approximately 75\% of good data), $\frac{W}{P}=0.1$, SNR=6, $\epsilon = 0.93$ \cite{FFA_2020MNRAS.497.4654M}, and $t_{\rm obs}=3600\ {\rm s}$, resulting in an approximate value of 2.4 $\rm \mu Jy$.
\begin{figure*}
	\centering
	\includegraphics[scale=0.48]{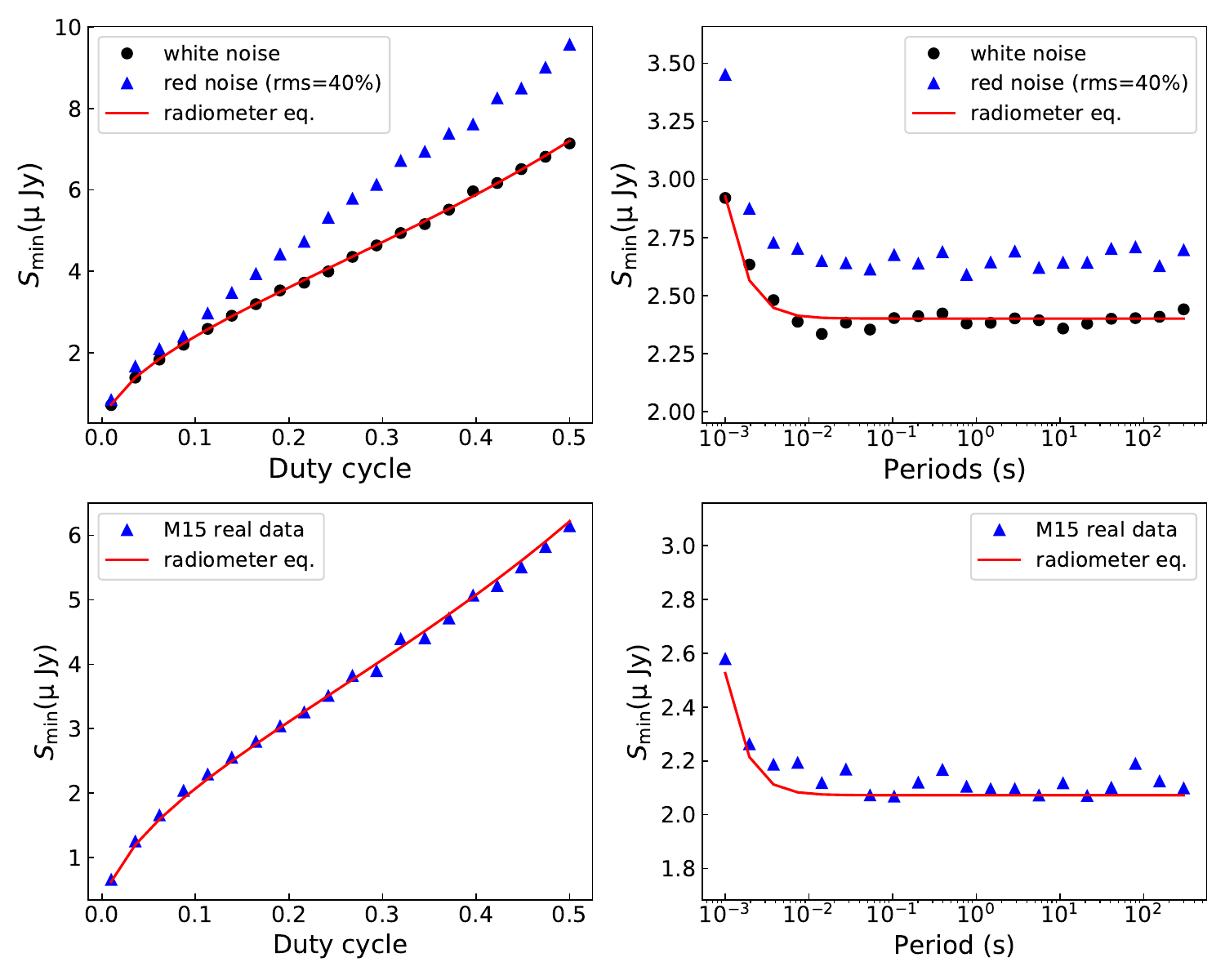}
	\caption{The simulated minimum detectable flux density of pulsars, using the algorithm mentioned in Section \ref{sec:red_smin}, as a function of pulse duty cycle and period. The top two panels depict the results obtained using simulated white noise and red noise with an rms of 40\%. The black circles correspond to the results for white noise, while the blue triangles correspond to the results for red noise. The red solid line represents the theoretical curve from the radiometer equation. It can be seen that the simulated values for the white noise case align with the theoretical values, while there is some deviation for the red noise case. The bottom two panels present the results obtained from the observational data of M15 observed by FAST in 2019. The blue triangles correspond to the computed results from the real M15 data, and the red solid line represents the theoretical curve from the radiometer equation. Despite the presence of red noise in the observational data, the deviation of the minimum detectable flux density from the theoretical values is not substantial, with a maximum deviation of approximately 10\% in the worst-case scenario.}
	\label{fig:simu_Smin}
\end{figure*}
\begin{figure*}
	\centering
	\includegraphics[scale=0.45]{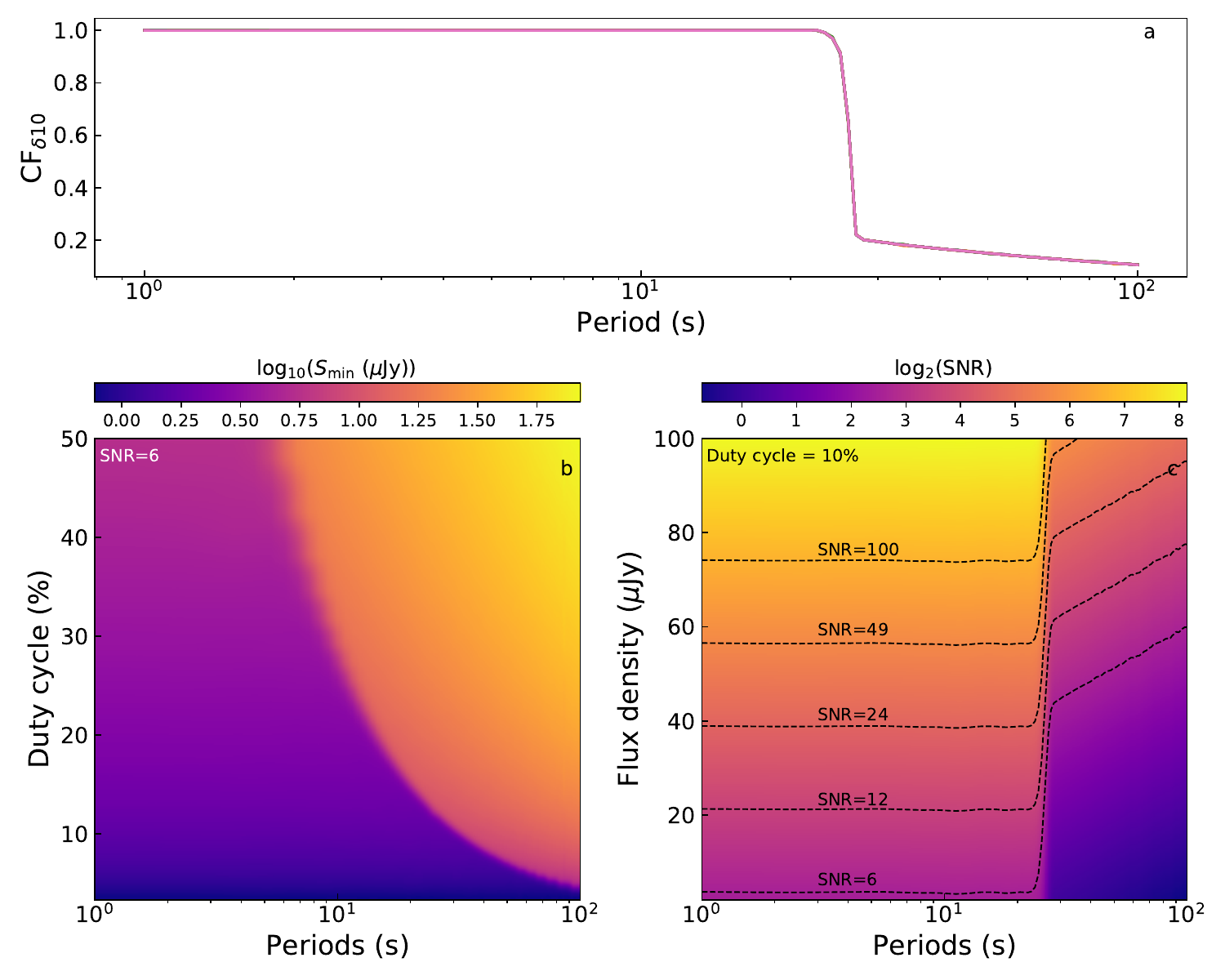}
	\caption{Panel a: Correction factors for the minimum detectable flux density at a 10\% duty cycle as a function of the signal period (see Section \ref{sec:red_smin} for details). Each curve represents the result of one observation, and these results overlap closely. Panel b: Heatmap of the minimum detectable flux density as a function of signal duty cycle and period when the SNR is 6. Panel c: Heatmap of SNR as a function of averaged flux density and period at a duty cycle of 10\%. The black dashed lines represent contours of the SNR.}
	\label{fig:GC_CF}
\end{figure*}

However, equation (\ref{eq:ro}) assumes that the data is dominated by white noise. In reality, with an increase in the telescope's integration time, variations in the telescope system temperature, sky background temperature, and gain changes will eventually introduce strong red noise into the data. The red noise disrupts the statistical assumption of equation (\ref{eq:ro}), rendering the calculation of the minimum detectable flux density of pulsars unreliable. Quantifying the specific impact of red noise on the minimum detectable flux density is exceptionally challenging. The study conducted by reference \cite{simulation_FFTg100_2015ApJ...812...81L} utilized simulated pulse injection to obtain the simulated variation of the minimum detectable flux density with pulse period. They found that the original formula significantly underestimated the minimum detectable flux density as the signal period increased. In our approach, we take a different route to obtain the actual minimum detectable flux density. We assume that the average temperature of the telescope remains nearly constant during the observation period, with only minor fluctuations around this average temperature. We then employ a multiple-sampling method to establish the relationship between telescope temperature fluctuations and the spectral characteristics of the measured time series, thus further establishing the connection between the time series and the minimum detectable flux density. The specific methodology is as follows:
\begin{itemize}
	\item Resample the input time series with an observation time of $t_{\rm obs}$ using a normal distribution. The center of the normal distribution is the measured value of the time series, and the standard deviation is taken as the measured value when the time sampling is 1 second (for other time samplings, it is converted according to the error synthesis formula). Then, obtain $m$ resampled time series.
	\item Calculate the mean value for each time series to obtain $m$ mean values.
	\item Calculate the standard deviation of these $m$ mean values, which serves as the measure of the telescope temperature fluctuation, denoted as $\Delta T$.
	\item For periodic pulse signals, assuming a duty cycle of $\delta$, the telescope temperature fluctuation still follows the following equation:
	\begin{align}
		\Delta T &= \sqrt{\Delta T_{\rm on}^2+\Delta T_{\rm off}^2}\\
		&= \sqrt{\Delta T(t_{\rm obs}=\delta T)^2+\Delta T(t_{\rm obs}=T-\delta T)^2}.
	\end{align}
	\item Assuming $T_{\rm peak}$ is the increase in telescope temperature caused by the pulsar signal, the ${\rm SNR}=\frac{T_{\rm peak}}{\Delta T}$ can be used to obtain $T_{\rm peak}$ under a given SNR using an iterative algorithm.
	\item Assuming the telescope gain is $G$, the minimum detectable flux density of the pulsar is $S_{\rm min}=\delta \frac{T_{\rm peak}}{G}$.
\end{itemize}

Using the above steps, the relationship between the time series and the minimum detectable flux density can be established without the need to ensure that the time series must be white noise. As a validation, we separately simulated white noise and red noise time series. Subsequently, we used our method to estimate the minimum detectable flux density of pulsars under these two time series and compared it with the radiometer equation. The simulation was performed using the \emph{simulate} module in \emph{stingray}\footnote{\url{https://github.com/StingraySoftware/stingray}} \cite{stingray_matteo_bachetti_2023_7970570}, a module based on the algorithm proposed by reference \cite{TK95_1995A&A...300..707T} for reverse simulating time series given a power spectrum density (PDS). By setting different root mean square (rms), we can adjust the strength of the simulated red noise, where smaller rms values make the simulated time series closer to white noise, and larger rms values result in stronger red noise. Fig. \ref{fig:simu_Smin} displays our simulation results. The top two panels of Fig \ref{fig:simu_Smin} depict curves of the simulated minimum detectable flux density as a function of duty cycle and period, considering both white noise and red noise with an rms of 40\%. These results are compared with theoretical values from the radiometer equation. It can be seen that the simulated values match well with the theoretical values in the case of white noise, but there is some deviation in the case of red noise. The bottom two panels of Fig. \ref{fig:simu_Smin} display results obtained from the observational data of the globular cluster M15 in 2019. Overall, the minimum detectable flux density shows minimal deviation from theoretical values, with a maximum difference of approximately 10\% in the worst-case scenario. This indicates that as long as the average temperature remains nearly constant, the differences in pulsar minimum detectable flux density caused by the red noise itself are negligible.

Although the minimum detectable flux density is not significantly impacted by red noise, not removing the red noise could lead to signal overlap, making it challenging to distinguish the signal. Although the signal is technically detectable once the red noise is removed, the removal process also eliminates signals on the same time scale, artificially reducing the signal amplitude and consequently increasing the actual minimum detectable flux density. In practical pulsar searches, especially for long-period pulsars (where, under constant duty cycle, the pulse width of the signal is larger and can be confused with red noise), it is necessary to eliminate red noise. This process increases the actual minimum detectable flux density, and the specific increase depends on the algorithm and parameters used for red noise removal. In our pulsar search, we employed a running median filtering algorithm to remove red noise. The use of running median filtering to eliminate red noise directly affects the signal amplitude reduction, leading to an increase in the average minimum flux density required for the original signal at a fixed SNR. The running median filtering algorithm's impact on pulse reduction is related to the pulse width; the algorithm preserves any pulse narrower than half the window width and weakens broader pulses \cite{medianfilter_1163708}. Therefore, it is evident that the increase in the minimum detectable flux density due to red noise removal is not directly related to the signal period but is directly related to the signal's time scale or pulse width. Assuming a nearly constant duty cycle for pulsar signals, longer signals will naturally have larger widths, making them more affected by red noise removal. To quantify this, assuming the duty cycle of pulsar signals is nearly constant and at a typical value of 0.1, and the running time window used for red noise removal is 5 seconds, red noise removal will significantly impact signals with periods longer than 25 seconds (5/2/0.1=25 seconds). The specific impact value may depend on the real data, so we use a simulation method to investigate the relationship between the actual minimum detectable flux density and the period after red noise removal. The specific approach involves injecting signals into each globular cluster dataset, then using running median filtering to remove red noise. By comparing the signal amplitudes before and after removal, we obtain the signal amplitude reduction ratio due to red noise removal, referred to as the correction factor ${\rm CF}_{\delta10}$. The correction factors for the globular cluster data we searched as a function of pulse period are shown in the panel a of Fig. \ref{fig:GC_CF}. It can be seen that, within a relatively short period range, the correction factor is generally close to 1, indicating that no correction is needed for the original minimum detectable flux density. However, for longer periods, the correction factor is noticeably less than 1, indicating the need for correction to the minimum detectable flux density. For a duty cycle of 10\%, the complete equation for the minimum detectable flux density is:
\begin{equation}
	S_{\rm min}=\frac{\beta(T_{\rm sys}+T_{\rm sky}){\rm SNR}}{{{\rm CF}_{\delta10}\epsilon G\sqrt{n_{\rm p}t_{\rm obs}\Delta f}}}\sqrt{\frac{W_{\delta10}}{P-W_{\delta10}}},
\end{equation}
where the dependence of the minimum detectable flux density on the period $P$ is partly included in ${\rm CF}_{\delta10}$. Since we assume that red noise has been eliminated, there will be no red noise factor in the expression. Panel b of Fig. \ref{fig:GC_CF} presents a heatmap of the minimum detectable pulsar flux density of the globular cluster data as a function of duty cycle and period after red noise removal, with an SNR of 6. It can be seen that, at any given duty cycle, the minimum detectable flux density increases with an increase in the period. However, the position where the minimum detectable flux density starts to increase is related to the duty cycle. This is because the running median filter is sensitive only to the pulse width of the pulsar signal, so as soon as the product of the duty cycle and the period reaches a threshold, the signal is rapidly attenuated. Panel c of Fig. \ref{fig:GC_CF} presents a heatmap of the SNR of pulsars in the globular cluster data as a function of the average pulsar flux density and period after red noise removal, with a duty cycle of 10\%. It can be seen that, for shorter periods, a relatively small average flux density is sufficient to achieve a high SNR. However, as the period increases, a larger average flux density is required to achieve the same SNR. Finally, we list the sensitivity of the pulsar search for each globular cluster data in Table \ref{tab:data}. As the minimum detectable flux density now depends on multiple parameters, we calculated values for periods of 1 second and 100 seconds at a duty cycle of 10\% and a SNR of 6 to obtain a typical range.

\subsection{Configuring search parameters}
From Fig. \ref{fig:GC_CF}, it can be seen that when the searched period exceeds 100 seconds, the correction factor is almost close to 0, making it extremely difficult to search for longer-period signals. While increasing the width of the running median filter can improve the correction factor for periods exceeding 100 seconds, it simultaneously makes the detection of signals at the second level difficult. Therefore, this is a balance selection problem. Considering that pulsars in globular clusters with periods at the second level are currently very rare, with only two reported \cite{long1_1993Natur.361...47L, long2_2022MNRAS.513.2292A}, we focus our search on the order of 100 milliseconds to 100 seconds. The width of the running median filter is set to 5 seconds.

As for the DM search, as mentioned in Section \ref{sec:FFA}, for globular clusters where pulsars have been discovered, a search is conducted within a range of $\pm 20 \ \rm pc \cdot cm^{-3}$ around their known DMs. For globular clusters where pulsars have not been discovered, the YMW16 model is used for DM estimation, and a search is conducted within a range of twice the estimated DM. From panel b of Fig. \ref{fig:GC_CF}, it can be seen that under high duty cycles and short periods, the simulated minimum detectable flux density based on the data is still acceptable. Therefore, to avoid missing signals with high duty cycles, we set the minimum pulse duty cycle for the search to 0.001 and the maximum to 0.5. Additionally, to ensure that long-period signals still have sufficient resolution at small duty cycles, we have appropriately adjusted the time resolution based on the size of the searched period.

\section{Discoveries}\label{sec:discoveries}
\begin{figure*}
	\centering
	\includegraphics[scale=0.5]{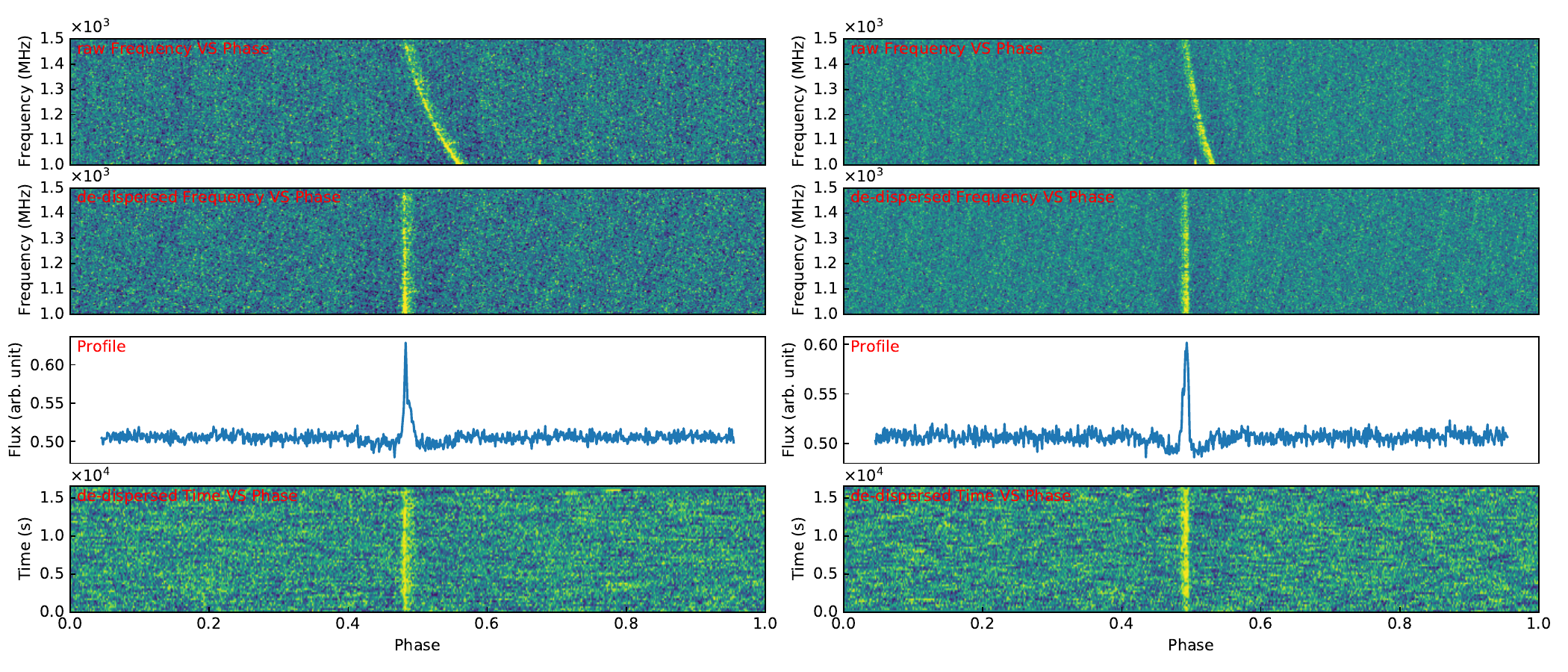}
	\caption{Long-period pulsars discovered in M15. The data used for the displayed figures is from MJD 59204.255287. Left panel: Long-period pulsar M15K with a period of approximately 1.928 seconds and a DM of about 66.5 $\rm pc\cdot cm^{-3}$. Right panel: Long-period pulsar M15L with a period of approximately 3.961 seconds and a DM of about 66.1 $\rm pc\cdot cm^{-3}$. In each subplot, from top to bottom, there are the frequency vs. phase plot without/with de-dispersion, the pulse profile, and the time-phase plot.}
	\label{fig: M15KL}
\end{figure*}
\begin{table*}
	\centering
	\caption{Observation information for 11 relatively long-duration observations of M15 observed by FAST.}
	\label{table:M15_11}
	\begin{tabular}{cc}
		\toprule
		Observation Date (UT) & Observation Length (s) \\
		\midrule
		2019/11/09 & 4800 \\
		2019/12/14 & 7320 \\
		2020/09/02 & 10680 \\
		2020/09/25 & 14400 \\
		2020/12/21 & 16200 \\
		2021/03/09 & 8700 \\
		2022/01/02 & 1800 \\
		2022/10/15 & 3000 \\
		2023/02/28 & 3000 \\
		2023/05/15 & 3000 \\
		2023/07/15 & 3000 \\
		\bottomrule
	\end{tabular}
\end{table*}
\begin{table*}
	\setlength\tabcolsep{10pt}
	\centering
	\captionsetup{justification=centering}
	\caption{Timing Solutions of Two Newly Discovered Long-period Pulsars in M15.}
	\label{tab:timing_fitresults}
	\renewcommand{\arraystretch}{1.0}
 \begin{threeparttable}
	\begin{tabular}{l c c} 
		\hline
		\hline
		Pulsar Name  &   M15K & M15L   \\
		\hline
		R.A. (J2000, h:m:s)   &   21:29:58.5(1)   & 21:29:57.92(9)\\
		DEC. (J2000, d:m:s)    &   +12:09:59(1)   & $+$12:09:34.4(9) \\
		Spin Frequency, $f$ (s$^{-1}$)   	&  0.51855097774(5) & 0.25247957712(2) \\
		1st Spin Frequency derivative, $\dot{f}$ (Hz\,s$^{-1}$)  &  $-$3.182(8)$\times 10^{-16}$ &$-$5.64(3)$\times 10^{-17}$\\
		Reference Epoch (MJD)   & 60000 & 60000\\
		Start of Timing Data (MJD)  & 58796.5 & 58796.5 \\
		End of Timing Data (MJD)  	& 60139.8 & 60139.8 \\
		Solar system ephemeris model\tnote{+}  & DE421 & DE421 \\
		Dispersion Measure (pc\,cm$^{-3}$) 	& 66.5(7) & 66.1(6)\\
		Number of ToAs      & 25 & 25\\
		Weighted rms residual ($\mu$s)   &  2804 & 3000 \\
		\hline
		\multicolumn{1}{c}{Derived Parameters} & M15K & M15L \\
		\hline
		Spin Period, $P$ (s)    & 1.9284507077(2) & 3.9607163930(3)\\ 
		1st Spin Period derivative, $\dot{P}$ (s\,s$^{-1}$)  &  1.183(3)$\times 10^{-15}$ & 8.85(5)$\times 10^{-16}$ \\
		Surface Magnetic Field, $B_0$ (G)  & 1.53$\times\ 10^{12}$ & 1.89$\times\ 10^{12}$\\
		Characteristic Age, $\tau_{\rm c}$ (Myr)  & 25.82 & 70.92 \\
            Position perpendicular offset from center\tnote{*} (arcmin/pc)  & 0.048/0.15  &  0.46/1.4 \\
		\hline
		\hline
	\end{tabular}
    \begin{tablenotes}
\item[+] TDB coordinate time system that has been used.
\item[*] The pulsar’s position is offset from the nominal cluster center ($21^{\rm h} 29^{\rm m} 58.33^{\rm s}$, $+12^{^{\circ}}10^{\prime}01.2^{\prime\prime}$) for a cluster distance 10.4 kpc (\url{https://physics.mcmaster.ca/~harris/mwgc.dat}).
 
    \end{tablenotes}
\end{threeparttable}
\end{table*}

\begin{figure*}
	\centering
	\includegraphics[scale=0.45]{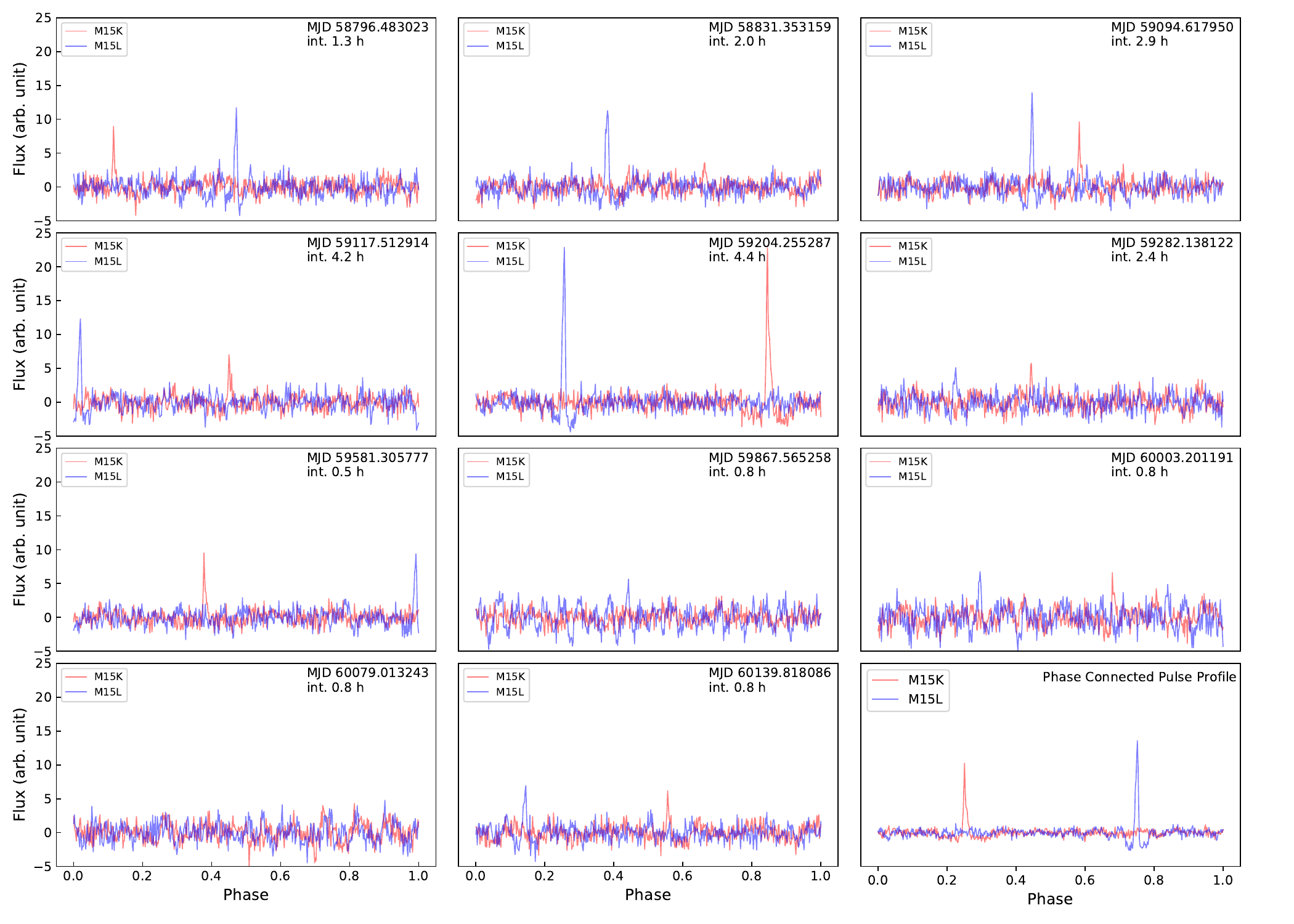}
	\caption{Pulse profiles for M15K and M15L obtained from FAST observational data listed in Table \ref{table:M15_11}. The data was filtered with a running median filter of width 0.3 seconds to mitigate red noise. The last panel displays the phase-connected pulse profile obtained through timing solutions (the phase has been appropriately shifted for ease of viewing). The varying SNR among different observations may be attributed to the potential effects of interstellar scintillation \cite{DanDan_2023SCPMA..6699511Z}.}
	\label{fig:M15KL_profile}
\end{figure*}

We systematically conducted a periodic search using FFA on the globular cluster data listed in Table \ref{tab:data}. Ultimately, we discovered two long-period pulsars in M15. The discovery of these two pulsars marks the first time that FAST has detected pulsars with periods on the order of seconds in globular clusters\footnote{\url{http://groups.bao.ac.cn/ism/CRAFTS/202310/t20231030_762168.html}}. Additionally, one of these pulsars has the longest period among pulsars discovered in globular clusters to date. We will introduce these two pulsars separately.

\subsection{M15K}
The pulsar M15K was discovered in the archived data from FAST on November 9, 2019 (UT). It's period is about 1.928 seconds, and the best DM is 66.5 $\rm pc\cdot cm^{-3}$. The frequency-phase plot, the pulse profile, and the time-phase plot, is shown in the left panel of Fig. \ref{fig: M15KL}. The pulse profile is single-peaked, with a duty cycle of approximately 1.09\% and an effective pulse width of 21.01 ms. We further searched the archived FAST data for additional observations of this globular cluster (details listed in Table \ref{table:M15_11}), and the pulse profiles for these observations are displayed in Fig. \ref{fig:M15KL_profile}. In most cases, the new search confirmed the existence of this pulsar. The lack of signals in some observations may be attributed to the potential effects of interstellar scintillation \cite{DanDan_2023SCPMA..6699511Z}.

The severe presence of red noise in the data significantly hindered the acquisition of TOAs. Therefore, we applied a running median filter with a width of 0.3 seconds to mitigate the red noise in the data. Finally, utilizing these archived data, and with the assistance of the code determining the rotation count of pulsars ("DRACULA" \cite{DRACULA_2018MNRAS.476.4794F}), we determined the phase-connected timing solution for M15K, spanning from 2019 to 2023. The timing solution is shown in Table \ref{tab:timing_fitresults} and the timing residuals are shown in the upper panel of Fig. \ref{fig:M15KL_timing_res}. According to the timing solution, the position of M15K relative to the center of M15 is shown in Fig. \ref{fig:M15KLcoord}. We can see that the position of M15K deviates by $2.9^{\prime\prime}$ from the center of the cluster, which is very close to the central location of the cluster.

\subsection{M15L}
The pulsar M15L was discovered in the archived data from FAST on November 9, 2019 (UT). It's period is about 3.961 seconds, and the best DM is 66.1 $\rm pc\cdot cm^{-3}$. The frequency-phase plot, the pulse profile, and the time-phase plot, is shown in the right panel of Fig. \ref{fig: M15KL}. The pulse profile is single-peaked, with a duty cycle of approximately 0.86\% and an effective pulse width of 33.93 ms. Similar to M15K, we further searched the archived FAST data for additional observations of this globular cluster (details listed in Table \ref{table:M15_11}), and the pulse profiles for these observations are displayed in Fig. \ref{fig:M15KL_profile}. In most cases, the new search confirmed the existence of this pulsar.

\begin{figure*}
	\centering
	\includegraphics[scale=0.7]{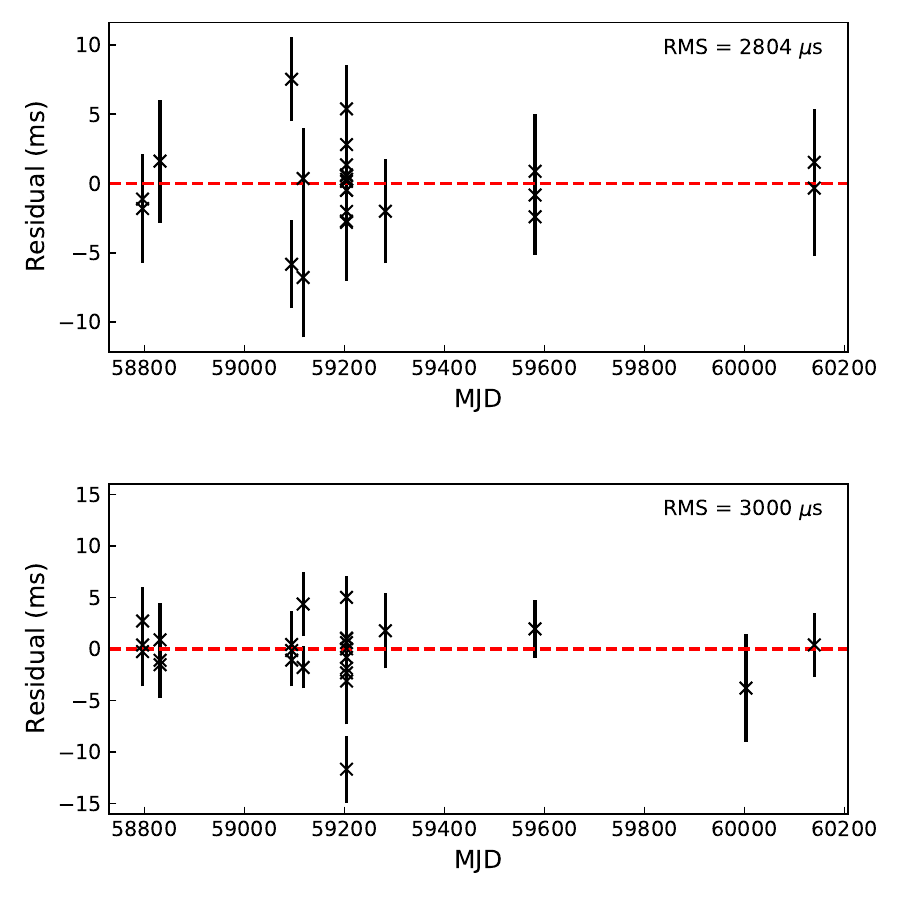}
	\caption{Upper panel: Timing residuals of M15K as a function of time. Lower panel: Timing residuals of M15L as a function of time.}
	\label{fig:M15KL_timing_res}
\end{figure*}

\begin{figure*}
	\centering
	\includegraphics[scale=0.8]{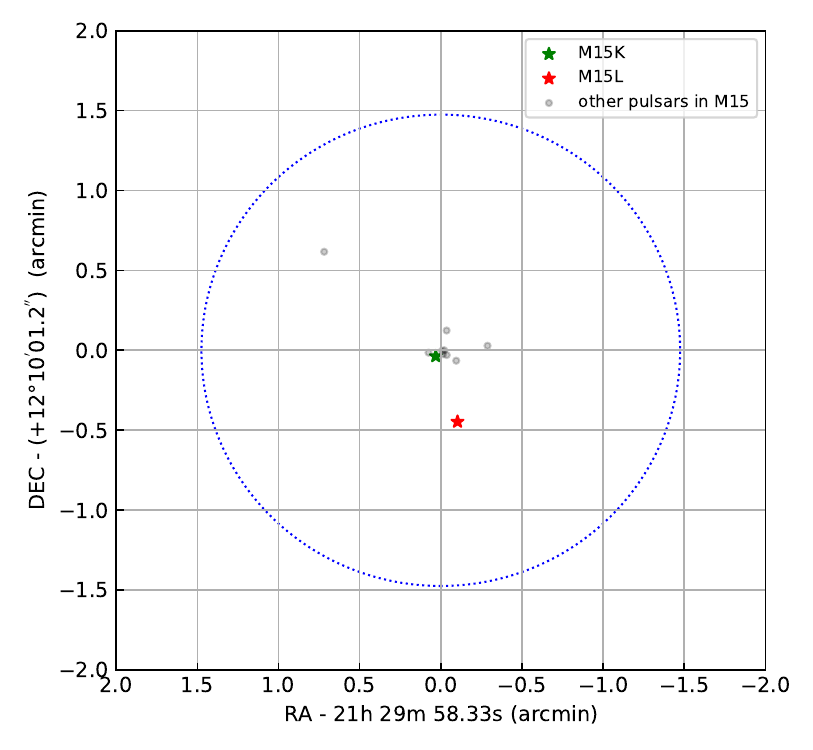}
	\caption{Map of all known and positioned pulsars in M15. The previously known pulsars are marked with gray dots, while the two newly discovered pulsars are shown in green and red pentagrams, respectively. The blue dashed circle represents the FWHM of a single beam of FAST ($\sim$ 2.95 arcmin).}
	\label{fig:M15KLcoord}
\end{figure*}

Consistent with the data processing approach for M15K, we applied a running median filter with a width of 0.3 seconds to mitigate the red noise in the data. Finally, these archived data enable us to determine the phase-connected timing solution for M15L that spans from 2019 to 2023. The timing solution is shown in Table \ref{tab:timing_fitresults} and the timing residuals are shown in the lower panel of Fig. \ref{fig:M15KL_timing_res}. According to the timing solution, the position of M15L relative to the center of M15 is shown in Fig. \ref{fig:M15KLcoord}. We can see that the position of M15L deviates by $27.5^{\prime\prime}$ from the center of the cluster but still falls within the cluster's half-mass radius of 1.06 arcmin. Unlike PSR J1823-3022 in NGC 6624 \cite{long2_2022MNRAS.513.2292A}, both the DM and position of M15L indicate that this pulsar resides in the globular cluster M15, marking it as the pulsar with the longest period discovered in globular clusters to date.
\begin{figure*}
	\centering
	\includegraphics[scale=0.28]{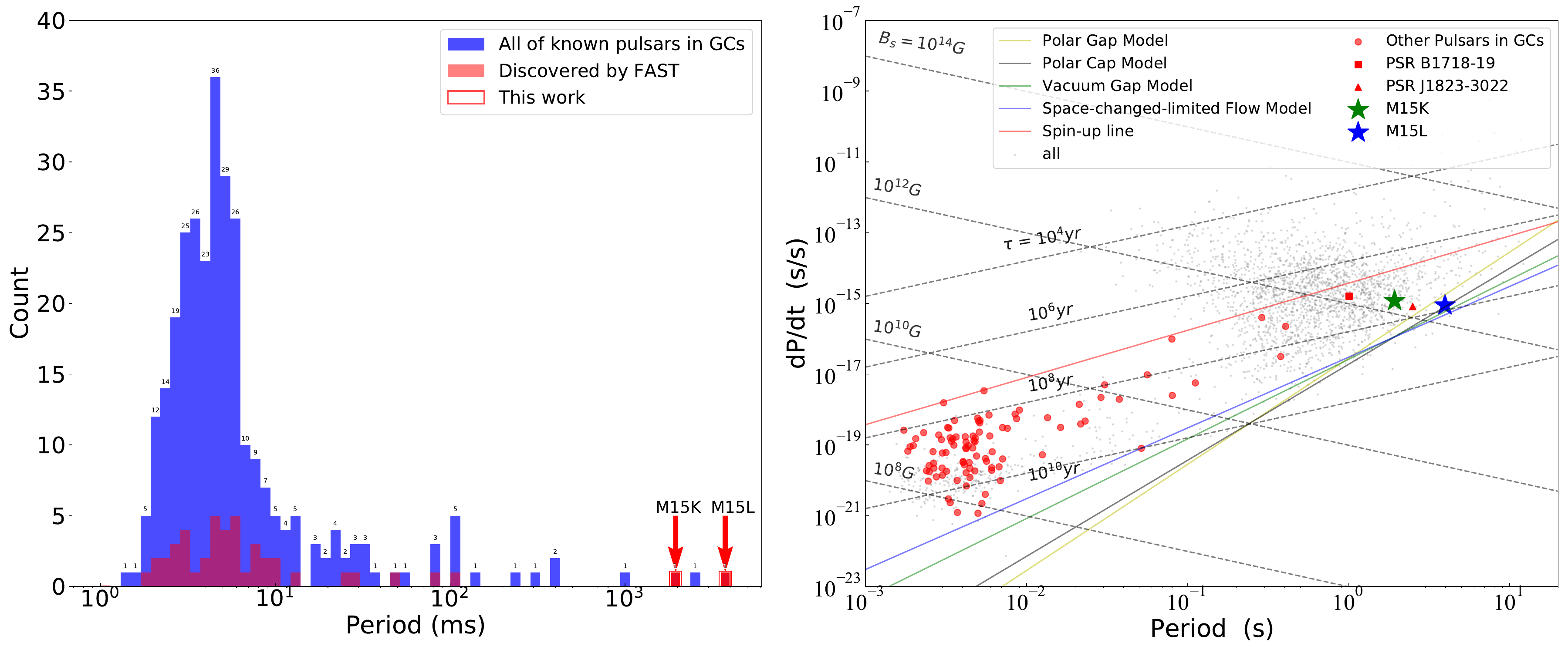}
	\caption{Left panel: the period distribution of pulsars in globular clusters. The blue histogram represents the period distribution for all pulsars in globular clusters, the red indicates those discovered by FAST, and the red-bordered ones are the pulsars discovered in this paper. It can be seen that the periods of the discovered pulsars are located at the edge of the period distribution for pulsars in globular clusters. Among them, the pulsar with a period of 3.928 seconds has the longest period among pulsars in globular clusters discovered to date. Right panel: The $P-\dot{P}$ diagram illustrates the measured period derivatives of M15K and M15L, marked with green and blue pentagrams, respectively. The PSR B1718-19 in the globular cluster NGC 6324 and PSR J1823-3022 in the globular cluster NGC 6624 are also highlighted with red rectangles and red triangles, respectively. Other pulsars in globular clusters are marked with red dots (most of them are MSPs), while all pulsars are shown against a background of gray dots. The plot displays lines of constant magnetic field, constant characteristic age, the spin-up line as adjusted by \protect\cite{explainSlow_2014A&A...561A..11V}, and four death lines obtained according to references \protect\cite{polar_gap_model}, \protect\cite{polar_cap_model} and \protect\cite{vacuum_gap_space-charged-limited_flow}.}
	\label{fig:pdis}
\end{figure*}

\section{Discussion}\label{sec:discussion}
The left panel of Fig. \ref{fig:pdis} shows the period distribution of all pulsars currently discovered in globular clusters. It can be seen that the majority of them are millisecond pulsars. The two pulsars, M15K and M15L, discovered in this study, are located at the outskirts of the distribution with long periods. 

The formation of long-period pulsars in globular clusters may be influenced by various factors. In the simplest scenario, isolated pulsars lose energy through radiation, eventually leading them to become long-period pulsars. In addition, dynamic interactions among celestial bodies, including tidal effects, collisions, and gravitational interactions \cite{tidal_1975MNRAS.172P..15F, MSP1_1991PhR...203....1B, explainSlow_2014A&A...561A..11V, long2_2022MNRAS.513.2292A,2023ApJ...944..225L, 2023MNRAS.525.4167O}, may also play a role in shaping the characteristics of these pulsars. In previous studies, two pulsars with periods in the order of seconds were detected in globular clusters, both associated with core-collapsed clusters. The first pulsar, PSR B1718-19, with a period of 1.004 seconds, is linked to the core-collapsed NGC 6342 cluster, exhibiting a substantial offset from the cluster center ($2.3^{\prime}$). This pulsar has a low-mass non-degenerate star as a companion, further strengthening its association with the cluster. The second pulsar, PSR J1823-3022, with a period of approximately 2.5 seconds, may be associated with the core-collapsed globular cluster NGC 6624. However, its larger offset from the cluster center ($3^{\prime}$) and noticeable deviation in DM from other pulsars in the cluster make its association less certain.

As for M15K and M15L, their association with the core-collapsed globular cluster M15 can be confirmed based on the positions obtained from the timing solutions and the typical DMs corresponding to the globular cluster. According to the timing solutions, we obtained the first period derivatives for M15K and M15L, which are $1.183\times 10^{-15}\ ss^{-1}$ and $8.85\times 10^{-16}\ ss^{-1}$, respectively. The corresponding surface magnetic field are $1.53\times 10^{12}\ {\rm G}$ and $1.89\times 10^{12}\ {\rm G}$, and the associated characteristic ages are 25.82 Myr and 70.92 Myr. Here, we provide a brief estimate of the impact of the acceleration due to the cluster's gravitational potential on the first derivative of the pulsar period. The derivative of the observed pulsar spin period is composed of the following factors \cite{1993ASPC...50..141P}:
\begin{equation}
    \left(\frac{\dot{P}}{P}\right)_{\rm obs} = \left(\frac{\dot{P}}{P}\right)_{\rm true}+\frac{\mathbf{a_p}\cdot \mathbf{n}}{c}+\frac{V_{\perp}^2}{cD},\label{eq:pdotp}
\end{equation}
where $\mathbf{a_p}$ is the acceleration of the pulsar, and $\mathbf{n}$ is the unit vector pointing from the earth to the pulsar. Therefore, $\mathbf{a_p}\cdot \mathbf{n}=\mathbf{a_l}$ is the line-of-sight acceleration. $V_{\perp}=\mu D$ is the pulsar's transverse velocity. $D$ is its distance from earth. The maximum value of $\mathbf{a_l}$ can be estimated by (\cite{1993ASPC...50..141P})
\begin{equation}
\text{max}|\mathbf{a_l}(R_{\perp})|=\frac{3}{2}\frac{\sigma^2}{\sqrt{r_c^2+R_{\perp}^2}},\label{eq:al}
\end{equation}
where $\sigma$ and $r_c$ are the values of the velocity dispersion at the center and the core radius of the globular cluster. For M15, $\sigma=13.5\rm\ km/s$, $r_c=0.424\rm\ pc$\footnote{\url{https://physics.mcmaster.ca/~harris/mwgc.dat}}. $R_{\perp}$ is the projected distance from the pulsar to the center of the cluster, and for M15K and M15L, they are equal to $0.15\rm\ pc$ and $1.4\rm\ pc$. For M15K and M15L, substituting the parameters into equation (\ref{eq:al}) yields values of ${\rm max}|\frac{\mathbf{a_p}\cdot \mathbf{n}}{c}|$ as $6.58\times 10^{-17}\rm\ s^{-1}$ and $2.02\times 10^{-17}\rm\ s^{-1}$, respectively. Using $D=10.4\rm\ kpc$\footnotemark[\value{footnote}] and $\mu=3.74\rm\ mas/year$ \cite{2006ApJ...644L.113J}, we find that the last term of equation (\ref{eq:pdotp}) is approximately $3.46\times 10^{-19}\rm\ s^{-1}$. The magnitudes of the period derivatives for M15K and M15L are both on the order of $10^{-15}\rm\ ss^{-1}$. Therefore, the maximum impact of acceleration on the $\dot{P}$ of M15K and M15L is only around 10\%, which will not change their orders of magnitude. The positions of M15K and M15L on the $P-\dot{P}$ diagram are shown in the right panel of Fig. \ref{fig:pdis}. We also plotted PSR B1718-19 and PSR J1823-3022 for comparison. We have also plotted the death lines of the four models \cite{polar_gap_model, polar_cap_model, vacuum_gap_space-charged-limited_flow} in the right panel Fig. \ref{fig:pdis} and find that M15L is relatively close to the death line. Previous theoretical models of pulsar radiative mechanisms have shown that cascade production is necessary to sustain the observed radio emission. The pulsar M15L in the region of the P-dot below the death line for the Polar Gap model is hard to support efficient pair cascade production above the pulsar polar caps in their inner magnetospheres. This is attributed to the unlikely increase in the thickness of the vacuum gap above the pulsar's polar cap over a long spin period, which is necessary to maintain the potential difference required for magnetic pair production. This leads to the disappearance of the radio emission. For another hand, both of M15K and M15L are above the death line of the space-charge-limited flow model, and if a multipolar magnetic field configuration is present, pair cascades can also be supported by the free flow of non-relativistic charges from the polar caps. The clear signature of a multipolar component near the surface of a neutron star can be seen in the magnetar SGR J0418+5729 \cite{Tiengo} and in a recent recycled MSP of PSR J0030+0045 \cite{Riley, 2019ApJ...887L..22R} suggesting that multipolar magnetic field configurations may be ubiquitous in neutron stars, including M15K and M15L.

The characteristic ages of these pulsars are not particularly old, especially for M15K. According to the assumption that young pulsars are recycled old netron stars, they should lie below the spin-up line on the $P-\dot{P}$ diagram \cite{explainSlow_2014A&A...561A..11V}:
\begin{equation}
	P_{\rm su} (s) \approx 1.6(10^{15}\dot{P})^{3/4}.
\end{equation}
This line represents the shortest spin period achievable through Eddington accretion during the recycling process. Following the suggestion of reference \cite{explainSlow_2014A&A...561A..11V}, we have plotted the spin-up line in the diagram, with the corresponding $\dot{P}$ values increased by a factor of 7 in order to account for the period derivatives of M28A amd NGC 6624A. It can be seen that both M15K and M15L are below this spin-up line, suggesting that these pulsars were partially recycled in an X-ray binary which was later disrupted by a dynamical encounter, as detailed in references \cite{explainSlow_2014A&A...561A..11V, 2023ApJ...944..225L, 2023MNRAS.525.4167O}. 

The magnetic field of M15K and M15L update the records of measured magnetic field for pulsars in globular clusters. If millisecond pulsars (MSPs) are formed during the accretion process, they may undergo a decay in magnetic field strength \cite{1974SvA....18..217B, 2011AIPC.1379...31B}. This results in the magnetic field of MSPs being in the range of $10^8\sim10^9\rm\ G$ \cite{2005AJ....129.1993M}. If pulsars are considered to undergo magnetic field decay during the accretion stage, then the strong magnetic fields of M15K and M15L ($\sim10^{12}\rm G$), which have only experienced partial accretion stages, can be considered understandable. This mechanism may be due to an increase in resistivity caused by heating, leading to a reduction in magnetic field strength \cite{1994MNRAS.271..490G}. Another proposal suggests that accreted material shields the magnetic field \cite{1990Natur.347..741R}, leading to magnetic field weakening during the accretion stage. Alternatively, reference \cite{2019MNRAS.490.2013C} propose that pulsars cool under constant magnetic fields before accretion, with an increase in rotation period. Subsequently, ambipolar diffusion causes magnetic field decay, and the rotation period approaches constancy. At this stage, neutron stars cross the death line, leading to the cessation of pulsar activity. During this period, pulsar temperature becomes very low, making it difficult to detect. When accretion occurs later, the neutron star's temperature rises, inhibiting ambipolar diffusion and magnetic field decay, causing an increase in rotation speed. Eventually, the neutron star transforms into a millisecond pulsar, returning above the death line, resulting in a short rotation period and a weak magnetic field. However, if M15K and M15L only undergo partial accretion, their magnetic fields should decrease due to ambipolar diffusion. Nevertheless, the inferred magnetic fields from their $P-\dot{P}$ relations are very strong ($\sim 10^{12}\rm\ G$), making this model challenging to explain the observed results.

The searched period range in this paper is from 0.1 seconds to 100 seconds. So far, the only pulsar with a period between 0.1 and 1 second in M15 is PSR B2127+11A (with a period of about 0.11 seconds), and we have also re-detected PSR B2127+11A in the data. PSR B2127+11A was first discovered by Wolszczan et al \cite{1989Natur.337..531W}. This pulsar is an isolated pulsar, and its period derivative is negative ($-2.107\times 10^{-17}\rm\ ss^{-1}$), which is explained as the result of the pulsar being bodily accelerated in our direction by the gravitational field of the collapsed core of M15. The long period of this pulsar is also considered to belong to the class of 'recycled' pulsars, which have been spun up by accretion in a binary system but later lost its companion, becoming isolated pulsar with gradually increasing period. This is similar to the newly discovered pulsars in our paper, implying that there may be more long-period isolated pulsars waiting to be discovered in M15.
\begin{table*}
	\centering
	\caption{The structural parameters of globular clusters studied in this paper.}  
	\label{tab:pars}
	\small
	\begin{threeparttable}
        \begin{tabular}{ccccccc}
                    \toprule
                    GC Name & $\Gamma_{\rm ss}$ &  $N_{\rm psr}$ & $\log{r_c}$ & $\log{\rho_c}$ & $\Gamma_{\rm sb}$\\
                            &   ($\Gamma_{\rm ssM15}$)      &      &     (pc)    & $(L_{\odot}\rm\ pc^{-3})$  &  ($\Gamma_{\rm sbM15}$) \\
                    \midrule
                    M53(NGC5024) & 0.008 & 3 & 0.26 & 3.07 & 0.024\\
                    NGC5053 & 0.000 & 0 & 1.02 & 0.54 & 0.000\\
                    M3(NGC5272) & 0.043 & 10 & 0.04 & 3.57 & 0.070\\
                    NGC5466 & 0.000 & 0 & 0.82 & 0.84 & 0.000\\
                    NGC5634 & 0.004 & 2 & -0.18 & 3.63 & 0.125\\
                    M5(NGC5904) & 0.036 & 9 & -0.02 & 3.88 & 0.115\\
                    NGC6229 & 0.011 & 4 & 0.03 & 3.54 & 0.070\\
                    M10(NGC6254) & 0.007 & 3 & -0.01 & 3.54 & 0.076\\
                    M92(NGC6341) & 0.060 & 13 & -0.20 & 4.30 & 0.285\\
                    M14(NGC6402) & 0.027 & 8 & 0.33 & 3.36 & 0.028\\
                    NGC6426 & 0.000 & 0 & 0.19 & 2.47 & 0.014\\
                    NGC6749 & 0.009 & 4 & 0.15 & 3.30 & 0.040\\
                    M56(NGC6779) & 0.006 & 3 & 0.08 & 3.28 & 0.046\\
                    M71(NGC6838) & 0.000 & 1 & -0.13 & 2.83 & 0.045\\
                    M72(NGC6981) & 0.001 & 1 & 0.36 & 2.38 & 0.009\\
                    M15(NGC7078) & 1.000 & 80 & -0.37 & 5.05 & 1.000\\
                    M2(NGC7089) & 0.115 & 20 & 0.03 & 4.00 & 0.118\\
                    \bottomrule
			\end{tabular}
		\begin{tablenotes}
			\item \textbf{Notes}. $r_c$, $\rho_c$ are obtained from \url{https://physics.mcmaster.ca/~harris/mwgc.dat}. $\Gamma_{ss}$ (normalized using the value of M15) and $N_{\rm psr}$ are taken from the reference \cite{2016RAA....16..151Z}. $\Gamma_{sb}$ is calculated using equation (\ref{eq:gsb}) and normalized using the value of M15.
		\end{tablenotes}
	\end{threeparttable}
\end{table*}

We discuss here briefly the implications of the newly discovered slow pulsars in terms of neutron star formation in globular clusters. The stellar encounter rate can be strongly correlated with the number of X-ray binaries as well as the number of radio pulsars in GCs \cite{explainSlow_2014A&A...561A..11V}. The discovery of M15K and M15L strengthens this scenario. The slow pulsars are found almost exclusively in clusters with a high encounter rate, consistent with these isolated slow pulsars being formed by the disruption of X-ray binaries, thus halting the recycling of a dead neutron star. This also implies that the higher the encounter rate in a specific globular cluster, the higher the number of isolated pulsars observed and the higher the number of previously missed detections of slow pulsars from another formation branch. It is noteworthy that, as shown in Table \ref{tab:data}, the sensitivity of M15 globular cluster with FAST is not outstanding. However, up to now, we have only discovered long-period pulsars in M15 globular cluster. References \cite{2010ApJ...714.1149H} and \cite{2013MNRAS.436.3720T} pointed out that in any given globular cluster, the number of potentially detectable pulsars $N_{\rm psr}$ mainly depends on the cluster's stellar encounter rate $\Gamma_{ss}$, with $\ln(N_{\rm psr})=-1.1+1.5\log \Gamma_{ss}$. Furthermore, since we are searching for long-period pulsars, and the formation of such pulsars in globular clusters is a result of the disruption of partially recycled pulsars, globular clusters with a high single-binary encounter rate ($\Gamma_{\rm sb}$) will host a greater number of isolated long-period pulsars, as indicated by reference \cite{explainSlow_2014A&A...561A..11V}. This parameter is calculated using the following equation:
\begin{equation}
    \Gamma_{\rm sb}\propto \frac{\sqrt{\rho_c}}{r_c},\label{eq:gsb}
\end{equation}
where $\rho_c$ and $r_c$ are the central luminosity density and the core radius of the globular cluster, respectively. In Table \ref{tab:pars}, we have presented the values of $\Gamma_{\rm ss}$, $N_{\rm psr}$, $r_c$, $\rho_c$, and $\Gamma_{\rm sb}$ for the globular clusters investigated in this study. From the table, it is evident that M15 has the highest values for $\Gamma_{\rm ss}$, $N_{\rm psr}$, and $\Gamma_{\rm sb}$. This indicates that, among the globular clusters studied in this paper, M15 has the highest probability of detecting pulsars, including long-period pulsars. Therefore, the fact that we have currently identified only two long-period pulsars in M15, with no such discoveries in other globular clusters, can be considered a normal phenomenon. Other globular clusters similar to M15 may also potentially harbor a population of missing long-period pulsars, and observations in X-rays hold the promise of aiding in the understanding of the intricate processes involved in the evolution of LMXB-MSP intermediate states. 

Another intriguing possibility is that these two pulsars have evolved from white dwarfs \cite{white_dwarf2023MNRAS.525L..22K}. The channels for the formation of isolated neutron stars include scenarios where a white dwarf exceeds the Chandrasekhar mass through accretion or when binary stars merge to form a neutron star. Therefore, if M15K and M15L are the result of the collapse of a white dwarf through accretion or the merger of two white dwarfs, and if these young pulsars, after undergoing prolonged evolution, are detected, then this explanation could be consistent with the interpretation of FRB20200120E in the globular cluster. This leads to the broader question of the origin of FRBs \cite{2022Natur.602..585K}.

\section{Summary}\label{sec:summary}
We systematically searched for long-period pulsars (100 ms $\sim$ 100 s) in globular clusters observed by FAST using FFA. Due to the presence of strong red noise in the data, it was necessary to perform red noise mitigation before searching for these pulsars. We proposed a simulation-based method to estimate the minimum detectable flux density of pulsars in the presence of red noise and compared the simulation results with radiometer equation. We found that the red noise in FAST has a minimal impact on the minimum detectable flux density. However, for separating pulsar signals from red noise, the removal of red noise is necessary, and this affects the minimum detectable flux density of pulsars. Red noise removal has a greater impact on signals with larger pulse widths. Therefore, for signals with a constant duty cycle, longer-period pulsars are more affected by red noise removal. We provided correction factors for the minimum detectable flux density due to red noise removal, calculated typical values of the corrected minimum detectable flux density for 1 s and 100 s, and listed them in Table \ref{tab:data}.

After the search, we discovered two long-period pulsars in the globular cluster M15, with periods of approximately 1.928 seconds and 3.961 seconds, respectively. Notably, the period of pulsar M15L surpasses even the longest known record among all pulsars in globular clusters, which is uncommon for clusters predominantly hosting millisecond pulsars. Additionally, the magnetic field derived from the $P-\dot{P}$ is exceptionally strong, whereas magnetic fields are generally considered to weaken during the pulsar accretion process. Furthermore, both M15K and M15L lie below the spin-up line in the $P-\dot{P}$ diagram. These pieces of evidence suggests that these pulsars may have undergone disruption during the recycling process, evolving into isolated pulsars and leading to a gradual increase in their periods. Moreover, another intriguing possibility is that these pulsars may have originated from accretion or mergers involving white dwarfs. The discovery of M15K and M15L provides a novel and significant sample for understanding this category of pulsars.

The positive results obtained by applying FFA instead of FFT for the search of long-period pulsars in globular clusters suggest that FFA indeed has a significant advantage in long-period pulsar searches. Future observations of globular clusters and iterative upgrades of algorithms may provide more opportunities to discover these faint, long-period pulsars. Once more long-period pulsars are found, it will contribute to a more comprehensive understanding of the evolutionary mechanisms of pulsars in globular clusters.
\Acknowledgements{We thank Prof. Yongfeng Huang from Nanjing University for the valuable discussion. This work was supported by the National Natural Science Foundation of China (Grant Nos. 11988101, 12103013, 12041303, U2031117, 12373109, 12103069), the National SKA Program of China (Grant Nos. 2020SKA0120200, 2022SKA0130100, 2022SKA0130104), the National Key R$\&$D Program of China (Grant No. 2023YFB4503300), the Foundation of Science and Technology of Guizhou Province (Grant No. (2021)023), the Foundation of Guizhou Provincial Education Department (Grant Nos. KY(2020)003, and KY(2023)059) and the ACAMAR Postdoctoral Fellowship. Di Li is a New Cornerstone Investigator. Pei Wang acknowledges support from the CAS Youth Interdisciplinary Team, the Youth Innovation Promotion Association CAS (id. 2021055), and the Cultivation Project for FAST Scientific Payoff and Research Achievement of CAMS-CAS. This work has used the data from the Five-hundred-meter Aperture Spherical radio Telescope (FAST). FAST is a Chinese national mega-science facility, operated by the National Astronomical Observatories of Chinese Academy of Sciences (NAOC).}

\InterestConflict{The authors declare that they have no conflict of interest.}


\bibliographystyle{scpma-zycai} 
\bibliography{ms}
\end{multicols}
\end{document}